\documentclass[runningheads]{llncs}
\usepackage[T1]{fontenc}
\usepackage[firstpage]{draftwatermark}
\usepackage{soul}
\usepackage{url}
\usepackage[hidelinks]{hyperref}
\usepackage[utf8]{inputenc}
\usepackage{tikz}
\usetikzlibrary{arrows.meta, positioning, fit, shapes}
\usepackage{graphicx}
\usepackage{color}

\usepackage{amsmath}
\usepackage{amssymb}
\usepackage{booktabs}
\usepackage{bm}
\usepackage{algorithm}
\usepackage{algorithmic}
\usepackage{bcprules}
\usepackage{chemarrow}
\usepackage{cite}

\usepackage{amsfonts}
\usepackage{xcolor}
\usepackage{ifthen}
\usepackage{nicefrac}
\usepackage{wrapfig}
\usepackage{dirtree}
\makeatletter
\let\MYcaption\@makecaption
\makeatother
\usepackage{subcaption}
\usepackage[normalem]{ulem}
\usepackage{wrapfig}
\usepackage{listings}
\allowdisplaybreaks[1]

\newcommand\Fig[1] {Fig.~\ref{#1}}

\newcommand\Sec[1] {Sect.~\ref{#1}}

\newcommand\Tbl[1] {Table~\ref{#1}}

\renewcommand{\phi} {\varphi}

\newcommand{\eqdef}{\ensuremath{\stackrel{\mathrm{def}}{=}}}

\newcommand{\Gospel}[0]{\textsf{Gospel}}
\newcommand{\WhyML}[0]{\text{WhyML}}

\newcommand{\WhyP}[0]{\textsf{Why3-py}}
\newcommand{\StatWhy}[0]{\textsf{StatWhy}}

\newcommand{\WhyT}[0]{\textsf{Why3}}

\newcommand{\cvcFive}[0]{\textsc{cvc5}}
\newcommand{\mypy}[0]{\text{mypy}}

\newcommand{\psiPre}[0]{\psi_{{\sf pre}}}

\newcommand{\phiPost}[0]{\phi_{{\sf post}}}

\newcommand{\alg}[0]{\mathit{A}}

\makeatletter
\providecommand{\leftsquigarrow}{%
  \mathrel{\mathpalette\reflect@squig\relax}%
}
\newcommand{\reflect@squig}[2]{%
  \reflectbox{$\m@th#1\rightsquigarrow$}%
}
\makeatother
\makeatletter
\newcommand{\subalign}[1]{%
  \vcenter{%
    \Let@ \restore@math@cr \default@tag
    \baselineskip\fontdimen10 \scriptfont\tw@
    \advance\baselineskip\fontdimen12 \scriptfont\tw@
    \lineskip\thr@@\fontdimen8 \scriptfont\thr@@
    \lineskiplimit\lineskip
    \ialign{\hfil$\m@th\scriptstyle##$&$\m@th\scriptstyle{}##$\hfil\crcr
      #1\crcr
    }%
  }%
}
\makeatother

\newcommand{\Know}{\mathbf{K}}

\newcommand{\KnowXx}[2]{\mathop{\mathbf{K}^{#1}_{#2}}}

\newcommand{\env}{\Gamma}

\newcommand{\triple}[3]{\{#1\}\ #2\ \{#3\}}

\newif\ifcommentson\commentsonfalse

\newif\ifconferenceon\conferenceonfalse
\ifconferenceon
\newcommand{\arxiv}[1]{}
\newcommand{\conference}[1]{#1}
\newcommand{\conferenceShort}[1]{}
\newcommand{\commentsize}[0]{.95\textwidth}
\else
\newcommand{\arxiv}[1]{#1}
\newcommand{\conference}[1]{}
\newcommand{\conferenceShort}[1]{}
\fi

\newif\ifanonymous\anonymousfalse

\ifcommentson
\newcommand{\commentYK}[1]{\begin{center} \parbox{\commentsize}{\textbf{\textcolor{black}{Comment Y.}} \textcolor{red}{#1} }\end{center}}

\newcommand{\replyYK}[1]{\begin{center} \parbox{\commentsize}{\textbf{Reply Y.} \textcolor{blue}{#1} }\end{center}}
\newcommand{\replyAT}[1]{\begin{center} \parbox{\commentsize}{\textbf{Reply A.} \textcolor{blue}{#1} }\end{center}}
\newcommand{\commentY}[1]{\marginpar{\footnotesize \color{red} {\bf Y:} \textsf{\scriptsize #1}}}
\newcommand{\commentAt}[1]{\marginpar{\footnotesize \color{red} {\bf AT:} \textsf{\scriptsize #1}}}
\newcommand{\replyY}[1]{\marginpar{\footnotesize \color{red} {\bf Y:} \textsf{\scriptsize #1}}}
\newcommand{\replyAt}[1]{\marginpar{\footnotesize \color{red} {\bf AT:} \textsf{\scriptsize #1}}}
\else
\newcommand{\commentYK}[1]{}
\newcommand{\commentAT}[1]{}
\newcommand{\replyYK}[1]{}
\newcommand{\replyAT}[1]{}
\newcommand{\commentY}[1]{}
\newcommand{\commentAt}[1]{}
\newcommand{\replyY}[1]{}
\newcommand{\replyAt}[1]{}
\fi

\definecolor{DarkGreen}{rgb}{0,.6,0}

\newcommand{\colorR}[1]{\textcolor{red}{#1}}

\newcommand{\commentOut}[1]{}

\newcommand{\pagelimitmarker}[1]{~\\ {\colorR{\ifthenelse{\thepage>#1}{\Huge Exceeding the page limit}{\huge Within the page limit}}}~\\ {\huge{\colorR{~~Page Limit\,\,\,\,\, = #1}}}~\\ {\huge{\colorR{~~Current Page = $\thepage$}}}}

\begin{document}
\title{{Why3-py}: A Tool for Formal Verification of Hypothesis Testing and Meta-Analysis in {Python}\thanks{The artifact of the paper is available at 
\texttt{\url{https://zenodo.org/records/21694782}} and 
\texttt{\url{https://github.com/fm4stats/why3-py}}.}
}
\titlerunning{Why3-py}
\ifanonymous
\author{Anonymized}
\institute{Anonymized}
\else
\author{Akira Tanaka\inst{1}
\orcidID{0000-0002-0820-0766} \and
Yusuke Kawamoto\inst{1}
\orcidID{0000-0002-2151-9560}
}
\authorrunning{A. Tanaka and Y. Kawamoto}
\institute{National Institute of Advanced Industrial Science and Technology (AIST), 
Tokyo, Japan
}
\maketitle              

\SetWatermarkText{\hspace*{3.7in}\raisebox{8.55in}{\href{https://eapls.org/pages/artifact_badges/}{\mbox{\includegraphics[width=1.4cm]{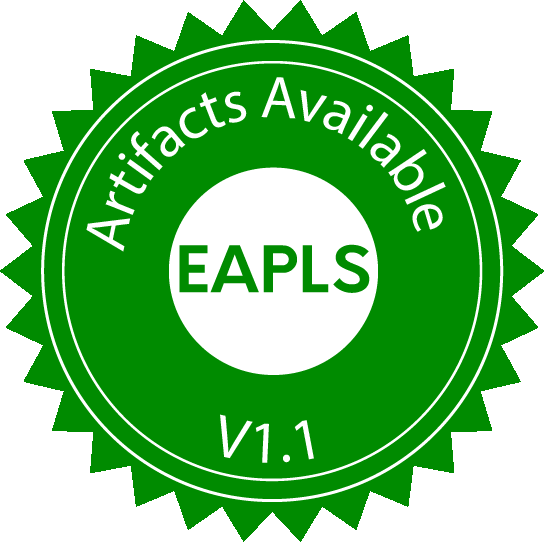}\includegraphics[width=1.4cm]{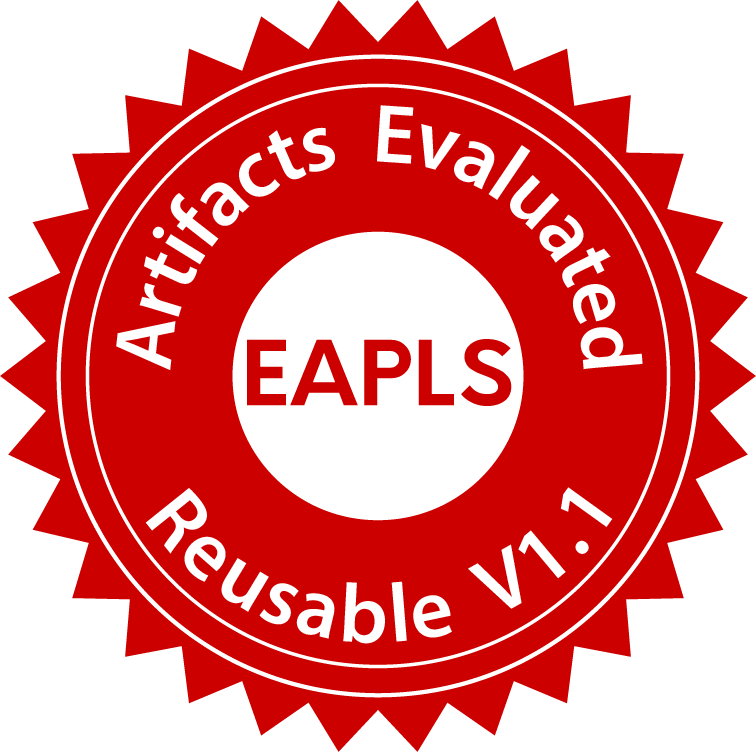}}}}}
\SetWatermarkAngle{0}
\begin{abstract}
The reproducibility crisis in scientific research has received widespread recognition, thereby increasing the importance of meta-analyses that integrate statistical analyses from multiple studies.
However, statistical methods often have ambiguous and implicit underlying assumptions, which can lead to their erroneous applications and interpretations.
To address this issue, we propose a formal verification framework for statistical Python programs.
Specifically, we present \WhyP{}, a Python front-end for the \WhyT{} verification platform that transforms Python code into verification-oriented WhyML representations, addressing the challenges arising from Python's dynamic typing and runtime polymorphism.
Furthermore, we extend the \StatWhy{} tool to support the verification of meta-analysis methods.
These tools enable meta-analysts to identify overlooked assumptions and misuse of analyses, and to verify the correct use of hypothesis testing and meta-analysis methods in Python code.

\keywords{Formal method \and automated verification \and program verification \and Why3 \and Python \and statistical hypothesis testing \and meta-analysis.}
\end{abstract}
\section{Introduction}
\label{sec:intro}
The reproducibility crisis in scientific research has drawn increasing attention to the trustworthiness of statistical analyses, particularly meta-analyses that aggregate evidence from multiple studies.
However, the correctness of statistical analyses is fundamentally different from that of conventional programs.
Statistical methods crucially rely on assumptions about data-generating processes, such as distributional properties of unobservable true populations, which cannot be verified solely from the program and observed data.
As a result, statistical programs may produce seemingly plausible outputs while being based on inappropriate or missing assumptions, leading to incorrect scientific conclusions.

This issue is exacerbated in practice by the widespread use of Python and its scientific libraries, which make it easy to apply statistical methods without explicitly stating their assumptions.
However, existing program verification techniques focus on functional correctness and do not capture the assumption-dependent nature of statistical reasoning or the usage of external libraries.

To address this gap,
we propose a formal verification framework for Python statistical code, based on explicit specifications of assumptions and interpretations.
Specifically, we present \WhyP{}, a Python front end for the \WhyT{} platform that transforms annotated Python code into \WhyML{}, and an extension of \StatWhy{}~\cite{Kawamoto:25:CAV} with meta-analysis specifications.
Importantly, our goal is not to verify implementations, but to ensure the correct use of statistical methods,
e.g., detecting incorrect integration of $p$-values in meta-analysis.
For this purpose, we redesigned \WhyT{}'s Python plugin to support real arithmetic and embed \WhyML{} specifications in Python code by 
developing a parser for mixed Python--\WhyML{} syntax and
incorporating static type information using \mypy{}~\cite{Lehtosalo:mypy}.

\noindent
\textbf{Contributions.}
Our main contributions are summarized as follows:
\begin{itemize}
\item We present \WhyP{}, a formal verification tool for Python statistical code.
\item We extend \StatWhy{} with specifications for meta-analysis methods and show how our approach detects missing assumptions and misuse of meta-analyses.
\end{itemize}
These tools
are available with a range of examples and documentation~\cite{Tanaka:26:manual:why3-py,Kawamoto:26:UD}.
To the best of our knowledge, this is the first approach 
that enables meta-analysts 
to formally verify Python statistical code.
We consider this work the first step toward developing a framework to verify the integrity of scientific conclusions.

\noindent
\textbf{Related Work.}
\noindent
\textit{Logic for Statistics.}
Modal logic has been employed to express statistical properties~\cite{Kawamoto:19:FC:short,Kawamoto:23:JELIA}.
The work on statistical epistemic logic~\cite{Kawamoto:19:FC:short,Kawamoto:19:SEFM,Kawamoto:20:SoSyM} is the first attempt to define a modal logic to describe properties of statistical methods.
\emph{Belief Hoare logic (BHL)}~\cite{Kawamoto:21:KR,Kawamoto:24:AIJ} is a program logic equipped with a modal operator for statistical belief.
Based on BHL, the \StatWhy{} tool~\cite{Kawamoto:25:CAV} automatically verifies 
the correct use of statistical methods in OCaml programs.

\noindent
\textit{Data Validation.}
A few works~\cite{Turcotte:25:FSE,Turcotte:25:ASE} have proposed methods for checking whether data satisfy statistical assumptions.
In contrast, \StatWhy{} with \WhyP{} focuses on formally proving the correct use of statistical methods in programs.

\noindent
\textit{Program Verification Tools.}
Deductive program verification aims to check a program's correctness by proving that the program satisfies its formal specification.
Various tools have been developed to statically verify the correctness of programs;
e.g., \WhyT{}~\cite{Filliatre:13:ESOP} for OCaml,
Dafny~\cite{DBLP:conf/lpar/Leino10} for imperative programs,
Frama-C~\cite{DBLP:conf/sefm/CuoqKKPSY12} for C programs,
and
KeY~\cite{KeYBook2016} for Java programs.

\noindent
\textit{Verification of Python Programs.}
Formal verification of Python programs remains relatively underexplored compared to statically typed languages, such as C, Java, and OCaml.
Nagini~\cite{Eilers:18:Nagini} provides automated verification for a statically typed subset of Python, based on the Viper intermediate verification infrastructure.
The PyVeritas tool translates Python code into C to apply bounded model checking~\cite{Orvalho:25:pyveritas}.
\WhyT{} provides a front end for micro-Python, a minimal subset of Python for education purposes that does not include real numbers or strings.
Several tools translate Python into Dafny~\cite{Leino:13:dafny,Li:25:dafny_intermediate, dafny_of_python_tool}.
However, these approaches 
require restrictive subsets of Python and precise semantic modeling.
In contrast, our approach supports practical Python statistical code by combining type reconstruction and abstraction of external libraries, enabling scalable verification.

\section{Background}
\label{sec:background}
\noindent
\textbf{Statistical Hypothesis Testing.}
\emph{Hypothesis testing}~\cite{Arbuthnot:1710} 
determines whether observed data provide sufficient evidence to support a claim called an \emph{alternative hypothesis}.
In a hypothesis test, the alternative hypothesis usually claims that there is an effect or relationship between variables, whereas the corresponding \emph{null hypothesis} represents a claim that there is no effect or no relationship.
The purpose of hypothesis testing is to assess whether the available evidence is sufficient to reject the null hypothesis in favor of the alternative hypothesis.
This assessment uses a \emph{$p$-value}, which quantifies how unlikely the observed data would have been generated under the null hypothesis.
The null hypothesis is rejected if the $p$-value is smaller than a threshold, e.g., $0.05$.

The correctness of a statistical program crucially depends on assumptions about the data-generating process; e.g., many hypothesis testing methods assume that the underlying population follows a normal distribution.
However, such assumptions are inherently unverifiable from the program and observed data, as they concern properties of an unknown true population which we can only partially and empirically learn from data.
Consequently, the correctness of statistical programs cannot be derived only from formally provable assertions.

\noindent
\textbf{Belief Hoare Logic (BHL).}
To address this limitation, \emph{belief Hoare logic} (\emph{BHL})~\cite{Kawamoto:21:KR,Kawamoto:24:AIJ} 
is designed as a program logic equipped with epistemic modal operators for the \emph{statistical beliefs} acquired by hypothesis testing.
The \emph{epistemic logic} used in BHL 
has two kinds of modal operators: $\Know$ and $\KnowXx{\le \epsilon}{y, A}$.
Intuitively, $\Know \phi$ represents that we know $\phi$.
The \emph{knowledge modality} $\Know$ is defined as the box operator in the S5 modal logic system.
Since the result of the hypothesis test may be wrong, we use the \emph{statistical belief modality} $\KnowXx{\le0.05}{y, A}$ instead of the knowledge modality $\Know$.
$\KnowXx{\le \epsilon}{y, A} \phi$ represents that by a hypothesis test $\alg$ on a dataset $y$, we believe $\phi$ with a $p$-value $\alpha \le \epsilon$.
The \emph{satisfaction} of a formula $\phi$ in a possible world $w$ is denoted by $w \models \phi$ and is defined straightforwardly in a Kripke model.

In BHL, we describe a procedure for hypothesis testing as a program $C$.
We then describe the requirements for the hypothesis tests as a \emph{precondition} formula $\psiPre$,
e.g., representing that a dataset has been sampled from a normal distribution.
We also specify the statistical belief acquired by the hypothesis test as a \emph{postcondition} formula $\phiPost$;
e.g., $\KnowXx{\le0.05}{y, A}\! \phi$
denotes that by a hypothesis test $\alg$ on a dataset $y$,
we believe an alternative hypothesis $\phi$ with a $p$-value $\alpha \le 0.05$.

Finally, we describe the whole inference as a \emph{judgment}
$\env\vdash\triple{ \psiPre }{ C }{ \phiPost }$,
representing that whenever the precondition $\psiPre$ is satisfied, the execution of the program $C$ results in the satisfaction of the postcondition $\phiPost$.
By deriving this judgment using derivation rules in BHL, we conclude that the program $C$ results in the statistical belief $\phiPost$ whenever the requirement $\psiPre$ is satisfied.

In this framework, specifications may include both verifiable properties and assumption-based statements.
This allows for a clear separation between (i) properties that can be formally verified through deductive reasoning and (ii) assumptions that must be justified empirically or externally.

\noindent
\textbf{Deductive Program Verification for Statistical Code.}
Using BHL as a theoretical foundation, \StatWhy{} provides a verification framework for statistical code.
Given an OCaml program annotated with a formal specification written in the \Gospel{} language~\cite{Chargueraud:19:FM}, \StatWhy{} verifies whether a programmer has appropriately annotated the program with the specification, i.e., the requirements for hypothesis testing and the interpretation of the test results.
\StatWhy{}'s implementation is based on the \WhyT{} verification platform~\cite{Filliatre:13:ESOP} 
to support compositional reasoning about programs with control flow and multiple statistical operations.

Importantly, this approach attempts not to prove the validity of statistical assumptions,
but to ensure that such assumptions are explicitly annotated, hence preventing the misuse and misinterpretation of statistical methods.
However, there remains a significant gap between the \WhyT{} platform and statistical programming practice, where Python and its rich libraries 
are widely used.

\section{Motivating Example: Meta-Analysis in Python}
\label{sec:overview}
\noindent
\textbf{Meta-Analysis of Harm Using Fisher's Method.}
Typical users of the \WhyP{} tool are meta-analysts who have a basic knowledge of statistical methods but wish to avoid mistakes in their application and interpretation.
To illustrate the main idea of \WhyP{}, consider a meta-analysis in which 
meta-analysts obtain multiple $p$-values, each corresponding to evidence from an independent study on the harm of a medicine, and then combine them into a single $p$-value that captures the accumulated evidence of harm using \emph{Fisher's method}.
In this setting, the goal of formal verification is not to prove the correctness of the implementation of Fisher's method, but rather to verify that the program uses Fisher's method appropriately to detect the harm of the medicine.

\begin{figure}[t]
  \vspace{-1.0em}
\begin{lstlisting}[frame = single, label=code1python]
def fail_meta_pvs_fisher(pv1 : real, pv2 : real, pv3 : real, observe_harm: formula) -> real :
  #@  requires \
  #@    let pvs = Cons pv1 (Cons pv2 (Cons pv3 Nil)) in \
  #@    pvalues pvs /\ intend_to_detect_excess_of_small_pv /\ \
  #@    forall w': world. (w' |= (Equiv indicate_risk (Impl disj_exps observe_harm))) \
  #@  ensures \
  #@    pvalue result /\ (World !st interp |= StatB (Eq result) indicate_risk)
  pvs = Cons(pv1, Cons(pv2, Cons(pv3, Nil)))
  return exec_combine_pvs_fisher(pvs, exps, disj_exps, observe_harm)

#@ execution
res = fail_meta_pvs_fisher(0.1, 0.2, 0.3, observe_harm)
\end{lstlisting}
  \vspace{-1.0em}
\caption{A Python program that uses Fisher's method to compute a meta-analytic $p$-value \texttt{res} from three independent studies with $p$-values \texttt{pv1}, \texttt{pv2}, and \texttt{pv3}.
\texttt{Cons} and \texttt{Nil} are used to construct a list of $p$-values corresponding to a \WhyML{} list, because a Python list is translated into a \WhyML{} array by \WhyP{}.
The formula \texttt{disj\_exps} is defined as the disjunction of the experimental situations \texttt{(sit exp1)}, \texttt{(sit exp2)}, and \texttt{(sit exp3)} outside the function.
\label{fig:illustrating:Python}}
\end{figure}

\noindent
\textbf{Structure of the Code.}
\Fig{fig:illustrating:Python}
shows a Python program that calls the function 
\texttt{exec\_combine\_pvs\_fisher},
which uses SciPy to compute a meta-analytic $p$-value \texttt{res} from three independent $p$-values \texttt{pv1}, \texttt{pv2}, and \texttt{pv3}
using Fisher's method.
This code consists of two intertwined layers.
The \emph{executable layer} is a Python program that calls the function \texttt{exec\_combine\_pvs\_fisher}.
The \emph{specification layer} is written in comments prefixed by \texttt{\#@}.
A programmer specifies the requirements and the interpretation of this function by annotating the code with the precondition in the \texttt{requires} clause and the postcondition in the \texttt{ensures} clause using \Gospel{}.
The \WhyP{} tool then parses these comments and generates VCs, which are discharged by \WhyT{}.
Thus, the same code serves as both an executable Python program and a formally specified verification target.
We remark that tool users need not develop formal specifications from scratch.
Instead, they can adapt the preconditions and postconditions provided by \StatWhy{}'s modules for individual statistical functions to their specific analyses.

\noindent
\textbf{Details of the Code.}
Although this example has a very simple program structure, its correctness depends on several assumptions, including non-trivial ones.
In the precondition, the formula $\texttt{pvalues pvs}$ checks that $\texttt{pvs}$ is a list of $p$-values, i.e., real numbers in $[0, 1]$.
$\texttt{intend\_to\_detect\_excess\_of\_small\_pv}$ indicates that the meta-analyst intends to 
detect an excess of small $p$-values, which is assumed for Fisher's method.
The formula $\texttt{indicate\_risk}$ specifies the analyst's decision rule, which stipulates that observing harm in one of the three experimental situations (denoted by \texttt{Impl disj\_exps observe\_harm}) leads to reporting a risk.
Finally, the postcondition states that the output $\texttt{result}$ is a $p$-value, and that by combining the three $p$-values using Fisher's method, we obtain a statistical belief that the alternative hypothesis 
$\texttt{indicate\_risk}$
holds 
with the combined $p$-value~$\texttt{res}$
in the current world $\texttt{(World !st interp)}$ equipped with a hypothesis testing record $\texttt{!st}$ and an interpretation $\texttt{interp}$ of private variables,
where $\texttt{!st}$ denotes the dereferenced value of the mutable reference $\texttt{st}$.

\begin{figure}[t]
\centering
  \includegraphics[width=\textwidth]{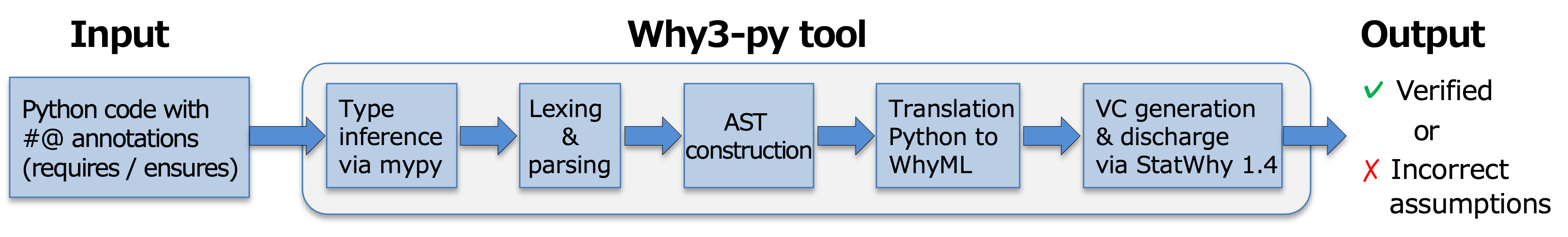}
  \vspace{-3ex}%
  \caption{Overview of the \WhyP{} verification workflow.}
  \label{fig:overview:why3-py}
\end{figure}

\noindent
\textbf{Program Verification and Error Detection.}
Given this Python code, the \WhyP{} tool verifies whether the meta-analysis method is appropriately used
by translating it into \WhyML{}, generating 
 \emph{verification conditions} (\emph{VCs})
using \WhyT{} and \StatWhy{} 1.4,
and discharging them using SMT solvers, e.g., \cvcFive{}.
Then \WhyP{} fails to verify this code owing to missing requirements in the annotation.
However, it helps identify the missing requirements by analyzing undischarged VCs.
For instance, analysts can easily identify the missing requirement that 
the three $p$-values should not be collected under \emph{publication bias} (i.e., the phenomenon that studies reporting statistically significant results are more likely to be published).
Importantly, such missing requirements do not manifest as runtime failures and are therefore difficult to detect in practice without 
verifying the specification annotated in the code.

\section{Extending \WhyT{} to Statistical Programs in Python}
\label{sec:pyton-VF}
\noindent
\textbf{Challenges in Verifying Python Statistical Programs.}
\WhyT{}'s micro-Python plugin has limited expressiveness and lacks support for real arithmetic and calls to external scientific libraries, which are crucial for statistical programming.
To address these limitations, we extend \WhyT{} to support the verification of Python programs that rely on statistical libraries, such as SciPy and statsmodels.
Our approach is based on the following key ideas.
First, we redesign the Python plugin to support a richer subset of Python and to integrate it seamlessly with \WhyT{}’s verification engine.
Second, we incorporate static type information obtained from a type checker to bridge the gap between Python’s dynamic typing and \WhyML{}, which is a statically typed language.
Third, we model statistical functions and external library calls using logical specifications, enabling the verification of their correct usage.
Finally, we extend the \StatWhy{} modules with meta-analysis specifications to integrate $p$-values from multiple studies.
Through these extensions, \WhyP{} enables analysts to verify the correct use of hypothesis testing and meta-analysis methods in Python programs.

\noindent
\textbf{A Python Front End for \WhyT{}.}
The \WhyP{} tool (\Fig{fig:overview:why3-py}) provides a new Python front end based on \WhyT{} 1.8.0 that targets a subset of Python programs that can be statically analyzed using \mypy{}~\cite{Lehtosalo:mypy}, a static type checker for Python.
Unlike the original micro-Python plugin, which relied on a plugin-specific, minimally expressive specification language, our tool supports writing logical specifications in the full \WhyML{} language within Python code.
This is achieved by embedding \WhyML{} expressions in comments of the Python code.
In addition, our Python front end supports real number arithmetic, which is essential for statistical programs.
It also translates Python-specific constructs, such as keyword arguments, into the appropriate \WhyML{} representations.
These design decisions are motivated by the need to support expressive constructs for statistical reasoning.
As a result, the new front end is not backward-compatible, but significantly improves the expressiveness of the specification for Python code.

\noindent
\textbf{Handling Dynamic Typing.}
Python's dynamic typing poses a significant challenge when translating programs into \WhyML{}, which requires explicit, static type information.
Our approach addresses this issue by using \mypy{} for static type checking and inference as a preprocessing step.
Although \mypy{} can infer the types of local variables from assignments, 
\WhyP{} still requires explicit type annotations for statistical functions in external libraries.
In our implementation, the verification pipeline first applies \mypy{} to the input Python code and uses the resulting type information to parse and construct the abstract syntax tree (AST).
This type information is essential for resolving ambiguities in Python syntax.
For instance, Python uses the same operator for both integer and real arithmetic, whereas \WhyML{} distinguishes between them.
Leveraging such type information, \WhyP{} translates Python operators into the appropriate \WhyML{} operators, ensuring the correct handling of numeric computations.
By restricting the input language to a subset of Python analyzable by \mypy{}, \WhyP{} enables sound and tractable verification while still supporting practical Python code.

\noindent
\textbf{Parsing a Mixed Python--\WhyML{} Language.}
To support the embedding of \WhyML{} specifications within Python code, we design a custom parser that handles a mixed Python--\WhyML{} language,
where lines prefixed with \texttt{\#@} are not ignored as Python comments, but are tokenized by lexers and parsed as \WhyML{} specifications.
Since Python and \WhyML{} have different lexical rules, 
parsing their combined syntax requires coordinating the two lexers according to the syntactic context maintained by the parser.
For this reason, \WhyP{} relies on Menhir’s incremental parsing API, which allows tokens to be supplied dynamically from the two lexers.
Specifically, the system maintains both a Python lexer and a \WhyML{} lexer, while using a single parser to process tokens from both lexers and managing input through a shared lexing buffer that stores the entire program in memory.
This enables the parser to backtrack and switch between the two lexers: the parser first invokes the Python lexer to interpret the input as Python.
If parsing fails, it resets the input position and invokes the \WhyML{} lexer to reinterpret the same fragment as \WhyML{}.
This design allows seamless switching between the two languages within a single source file.

\noindent
\textbf{Translation to \WhyML{}.}
After parsing, the constructed Python AST is translated into a \WhyML{} AST.
In this translation, 
Python function calls are mapped to corresponding \WhyML{} function calls, and keyword arguments are translated into record constructions.
Python lists are translated into \WhyML{} arrays, whereas lists constructed using \texttt{Cons} and \texttt{Nil} are translated into \WhyML{} lists.
Logical expressions embedded in Python are represented as \WhyML{} terms.

\noindent
\textbf{Verification with External Libraries.}
External libraries, such as SciPy, cannot be directly verified using \WhyT{}.
To address this limitation, we adopt a specification-based abstraction approach in \WhyP{}.
Instead of modeling the implementations of external functions, we associate them with logical contracts that formalize their behavior in our \StatWhy{} 1.4 modules.
Verification is then performed under the assumption that these contracts are satisfied.
This approach allows us to reason about the correct use of library functions, rather than their internal implementation details.
As a result, \WhyP{} supports modular verification of Python statistical code that depends on external scientific libraries.

\noindent
\textbf{Scope of Supported Python Code.}
\WhyP{} targets a subset of Python that can be statically analyzed using \mypy{}.
This includes numerical computations, conditional branching, loops, and calls to external libraries equipped with logical specifications.
However, programs relying heavily on dynamic features such as dynamically constructed data structures are currently outside the scope.
\conference{%
The details of the syntax are presented in Appendix~\ref{sec:syntax:WhyP}.
}%

\section{Case Studies and Performance Evaluation}
\label{sec:applications}

\begin{table}[t]
    \vspace{-1.0em}
    \centering
    \caption[multiple-comparison-methods]{
    Verification times (in seconds) for popular multiple comparison methods covered in standard textbooks.
    \#groups (resp. \#comparisons) denotes the practical number of groups (resp. combinations of groups) compared in the tests.
    }
    \label{tab:multiple-comparison-methods}
    {
    \tabcolsep=4pt
    \renewcommand{\arraystretch}{1}
    \begin{small}
    \begin{tabular}{llrrrrrr}
    \toprule
    & & \multicolumn{6}{c}{\#groups} \\
    \cmidrule(lr){3-8}
    \multicolumn{1}{c}{Test methods} & \multicolumn{1}{c}{Metric} & \multicolumn{1}{c}{2} & \multicolumn{1}{c}{3} & \multicolumn{1}{c}{4} & \multicolumn{1}{c}{5} & \multicolumn{1}{c}{6} & \multicolumn{1}{c}{7} \\
    \midrule
    Tukey's HSD test & Time (sec)      & 10.90 & 20.73 & 31.95 & 47.65 & 65.97 & 87.86 \\
                     & \#comparisons & 1    & 3    & 6    & 10   & 15    & 21 \\
    Dunnett's test   & Time (sec)      & 10.03 & 18.71 & 22.51 & 26.71 & 32.90  & 35.39 \\
                     & \#comparisons & 1    & 2    & 3    & 4    & 5     & 6 \\
    Steel-Dwass' test & Time (sec)     & 10.61 & 20.30 & 31.32 & 46.05 & 65.60 & 93.72 \\
                     & \#comparisons & 1    & 3    & 6    & 10   & 15    & 21 \\
    \bottomrule
    \end{tabular}
    \end{small}
    }
\end{table}

\noindent
\textbf{Detecting Incorrect Requirements for Meta-Analyses.}
Our \StatWhy{} 1.4 supports the \WhyML{} specification of various meta-analysis methods, including \emph{Fisher's method}, \emph{Stouffer's method}, and \emph{Mantel-Haenszel's method}.
Relying on them, \WhyP{} can verify Python meta-analysis code including a specification (around 20 to 40 AST nodes) and detect various incorrect requirements.
One of the important requirements is the absence of \emph{publication bias}, as discussed in \Sec{sec:overview}.
If meta-analysts do not annotate the code with the requirement \texttt{sampled d uniform\_pv}, then \WhyP{} detects that this requirement is missing,
indicating that meta-analysts cannot reliably apply Fisher's method under publication bias.

\noindent
\textbf{Detecting $P$-hacking in Multiple Comparisons.}
\WhyP{} can also be applied to Python hypothesis testing code, including multiple comparisons within a single study.
Such settings may involve \emph{$p$-hacking}, where multiple hypotheses are tested simultaneously and $p$-values are selectively reported or manipulated, which can lead to incorrect scientific conclusions.
By annotating Python code with its specification (around 30 AST nodes) and verifying it using \WhyP{}, analysts can detect incorrect requirements and interpretations of $p$-values.

\noindent
\textbf{Scalability of \WhyP{}.}
We evaluated the verification performance of Python programs of approximately 40 lines of code each using \WhyP{} and \StatWhy{} 1.4, varying  
(i) the number of $p$-values integrated in meta-analyses and (ii) the number of groups compared in multiple comparisons.
Our experimental evaluation uses
the SMT solver \cvcFive{} 1.1.2
on a ThinkPad 480s with Debian 13, an Intel Core i7-8650U, and 24 GB of memory.
The verification times for Python programs computing $p$-values using each meta-analysis method are within 8 seconds for practical numbers (e.g., 2 to 30) of studies.
\conference{(See \Tbl{tab:meta-analyses} in Appendix~\ref{sub:performance:meta}.)}
The verification times for 
the Python 
code for
multiple comparisons
are shown in \Tbl{tab:multiple-comparison-methods}.
Although the number of comparisons between groups grows rapidly with the number of groups,
the verification of these programs is efficient for practical numbers of groups.

\section{Conclusion}
\label{sec:conclude}
We proposed a formal verification framework for Python statistical code.
Specifically, we developed the verification tool \WhyP{} for Python statistical code
and extended \StatWhy{} to specify meta-analyses.
To the best of our knowledge, this is the first approach to formally verifying Python statistical code.

\begin{credits}
\subsubsection{\ackname} 
The authors were supported by JSPS KAKENHI Grant Number JP24K02924, Japan.
Yusuke Kawamoto was supported by JST FOREST Grant Number JPMJFR242M, Japan.
\end{credits}
\bibliographystyle{splncs04}
\bibliography{short,short-stat,sv,python}

\end{document}